\input{psfig}
\documentclass[referee]{aa}
\usepackage{astrobib,graphicx}
\input psfig.sty
\newcommand{\be}{\begin{equation}}
\newcommand{\ee}{\end{equation}}
\newcommand{\bea}{\begin{eqnarray}}
\newcommand{\eea}{\end{eqnarray}}
\begin{document}
\title{Orbital migration and the period distribution of exoplanets}

\author{A. Del Popolo\inst{1,2}, N. Ercan\inst{1}, \& I.S. Ye\c{s}ilyurt\inst{1}}
\titlerunning{Orbital migration}
\authorrunning{A. Del Popolo et al.}
\date{}
\offprints{A. Del Popolo, E-mail:antonino.delpopolo@boun.edu.tr}
\institute{
$1$  Bo$\breve{g}azi$\c{c}i University, Physics Department,
     80815 Bebek, Istanbul, Turkey\\
$^2$ Dipartimento di Matematica, Universit\`{a} Statale di Bergamo,
  via dei Caniana, 2,  24127, Bergamo, ITALY 
}
\maketitle


\maketitle

\begin{abstract}
We use the model for the migration of
planets introduced in Del Popolo, Yesilyurt \& Ercan (2003)  
to calculate the observed mass and semimajor axis distribution of extra-solar planets.
The assumption that the surface density in planetesimals is
proportional to that of gas is relaxed, and in order to describe disc evolution 
we use a method 
which, using a series of simplifying assumptions, is able
to simultaneously follow the evolution of gas and solid particles for
up to $10^7 {\rm yr}$. 
The distribution of planetesimals
obtained after $10^7 {\rm yr}$ is used to study the migration rate of
a giant planet through the model of this paper.
The disk and migration models are used to calculate the distribution 
of planets as function of mass and semimajor axis. The results show that the
model can give a reasonable prediction of planets' semi-major axes and mass distribution.
In particular there is a pile-up of planets at $a \simeq 0.05$ AU, a minimum near 0.3 AU, 
indicating a paucity of planets at that distance, and a  rise
for semi-major axes larger than 0.3 AU, out to 3 AU. 
The semi-major axis distribution shows that the more massive planets (typically, masses larger
than $4 M_{\rm J}$) form preferentially in the outer regions 
and do not migrate much. Intermediate-mass objects migrate more easily
whatever the distance they form, and that the lighter planets (masses from sub-Saturnian to
Jovian) migrate easily.

\end{abstract}


\section{Introduction}

The number of currently known extra-solar planets exceeds 100. This is a 
statistically significant sample from which it is possible to determine meaningful 
distributions of planetary characteristics (Udry et al. 2003, Marcy et al. 2003), and  useful constraints for the planet-building models, and then possibly 
discriminate between the different proposed scenarios. 
Some features in the period
versus mass diagram start to emerge and provide strong
observational constraints for the migration scenario. 

After the discovery of solar-like stars showing evidences of planets
orbiting around them (\citeNP{Mayor1}, \citeNP{Marcy1},
\citeNP{Vogt1}, \citeNP{Butler1}) 
a number of
questions about the formation mechanisms of such systems have been raised. 
The properties of these planets, most of which are Jupiter-mass objects \footnote{Note that the Doppler 
method measures only M sini, and in the following when we refer to planet masses, we mean M sini },
and with orbits ($r< 1 {\rm AU}$) very close to their stars, 
are difficult to explain using the standard model for planet formation 
(Lissauer 1993; Boss 1995),
which 
predicts nearly circular planetary orbits and giant
planet distances $\geq 1$ AU from the central star, where the temperature 
in the protostellar nebula is low enough for icy materials to condense 
(Boss 1995; Boss 1996; but see also Wuchterl 1993;
Wuchterl 1996). 
Thus, in the case of close-in giants, it is very unlikely that such planets 
were formed at their present locations: the most 
natural explanation for this paradox and for planets on very short orbits is
that these planets formed further away in the protoplanetary
nebula and migrated afterwards to the small orbital distances at
which they are observed (\citeNP{Rasio1}; \citeNP{Weid1}; \citeNP{Murray1};
\citeNP{Gold1}; \citeNP{Gold2}; \citeNP{Ward1}; 
\citeNP{Lin4}; \citeNP{Ward2}; Del Popolo, Gambera and Ercan 2001 (hereafter DP1)).
Recent studies based upon the quoted sample have shown some peculiar 
features that could be used to constraint theoretical models (migration models), 
and have generated new puzzles that should be solved. \footnote{In the list of new problems, a noteworthy role can be given to the origin of eccentricities 
for single planets and the steep planet mass distribution. Plausible mechanisms include gravitational 
interactions between the planet and other planets, stellar companion, or the protoplanetary disk (Rasio \& Ford 1996; 
Lin \& Ida 1997; Marzari \& Weidenschilling 2002; Goldreich \& Sari 2002; Chiang 2003).
The dynamical processes that lead to resonances remain under study.}
Some of the emerging observational properties are the following (Udry et al. 2003):\\
1) The more massive planets (typically, masses larger
than $4 M_{J}$) form preferentially in the outer regions 
where there is a large-enough material reservoir and do
not migrate much. None are observed within 
($\simeq$ 100 days) 
of the central star.\\
2) The clear emergence of a gap in the 10 and 100-days periods (the ``period valley" following Udry et al. 2003), 
 due to the light candidates
($M_{\rm p} \sin{i} \leq 2 M_{\rm J}$).\\
3) The apparent lack of very light planets ($M_{\rm p} \sin{i} \leq 0.75 M_{\rm J}$) with 
periods larger than $\simeq 100$ days.
Most of them are 
found in the very central regions. None is observed
with $P \ge ~100$ days. 
\\

The semi-major axes distribution shows:\\
i) A pile-up of planets at $a \simeq 0.05$ AU. \\ 
ii)  A  minimum near 0.3 AU.\\
%
%

The different features pointed out above in the 
observed mass versus period and in semi-major axes distribution can be used to check
the processes proposed to explain the observed distribution of mass and periods for extra-solar planet candidates, and
fix some constraints for the planet migration models as a function of the planet mass, in single star systems.

Some studies have tried to reproduce the quoted behaviors. 
The lack of massive planets on short-period orbits can be explained by means of a low migration efficiency
for higher mass planets. 
When the mass of the planet becomes of the order of a significant fraction
of the characteristic disc mass with which it interacts, the
inertia of the planet becomes important and slows down
the orbital evolution (e.g. Trilling et al. 1998; Nelson et al.
2000). 
Using a model assuming a simple impulse approximation
for the type II migration, no ad-hoc stopping mechanism
in the center and neglecting type I migration,
Trilling et al. (2002) showed that, for identical initial disk
conditions, close-in surviving pseudo-planets have smaller
masses and distant surviving pseudo-planets have larger
masses. 
Armitage et al. (2002) extended the work of Trilling et al. 
(2002) including a physical mechanism for disc 
dispersal into a model for the formation and migration of 
massive planets, and made quantitative comparisons with the 
observed distribution of planetary orbital radii. 
They found good agreement with the Trilling work.

A recent study by Masset \& Papaloizou (2003) gives
some insight into 
the observed lack of
small-mass planets ($M_{\rm p} \sin{i} \leq 0.75 M_{\rm J}$) on intermediate-period
orbits ($P \ge 100$ days) (see the Results section of the present paper, for details).

The aim of the present paper is to check the predictions of the migration model introduced in 
DP1, for the semi-major axes and mass distribution 
of extra-solar planets, and to compare them with the observed distribution of 
known extra-solar planets. 
This goal is reached by allowing giant planets to evolve and migrate in 
circumstellar disks with various initial conditions (different values of planet mass,
disk mass, viscosity, etc.), assuming one planet per disk, and determining the final semi-major axis and mass
distributions.
The fundamental differences to previous studies in literature lie in the 
disk model used and the migration model.

A fundamental issue in studying planet migration is the disk model. 
From Del Popolo, Gambera and Ercan (2001) to 
Del Popolo, Yesilyurt and Ercan (2003) (hereafter DP3), 
there was a noteworthy improvement in the disk model used. 
In the current work, as in DP3, the disk model we used is a
time-dependent accretion disk, 
and moreover we relaxed the 
assumption of Del Popolo \& Ek\c{s}i (2002) (hereafter DP2) 
that the surface density in planetesimals remains proportional to that of gas:
$\Sigma_{\rm s}(r,t) \propto \Sigma(r,t)$. It is well-known
that the distribution of planetesimals emerging from a turbulent disk
does not necessarily reflect that of gas (e.g., \citeNP{SV1}, \citeNP{SV2}). 

In order to take into account that the evolution of the distribution 
of solids is also governed by coagulation, sedimentation and evaporation/condensation, 
we use the method developed in
Stepinski \& Valageas (1997) which is able to simultaneously follow the evolution of
gas and solid particles for up to $10^7 {\rm yr}$. 

The second fundamental issue in studying planet migration is the migration mechanism. 
A detailed discussion of the mechanisms that 
have been proposed to explain the presence of planets at small orbital 
distances are summarized in DP1 and  DP2, and in those papers we showed that dynamical friction
between a planet and a planetesimal disk is an important mechanism
for planet migration.
Coupling the two models to obtain the migration rate of 
planets is possible, as shown in DP1, DP2, and DP3,  and this can be used to calculate the distribution in terms of semimajor axes and mass. 
%
%

This paper is organized as follows. In Sect. 2, we revise the disk and migration model used to
obtain the radial distribution of the planetesimals. 
In Sect. 3, we describe our results. Sect. 4 is devoted to conclusions.

\section{Disk and migration models}
\subsection{Distribution and evolution of gas and solids}
%
In this section, we summarize the models used for the disk and migration. Interested readers are referred to DP3, and previous papers of the same author, for more details.

In order to obtain a better description of the planets' migration, 
it is necessary to model how the 
structure of the disk changes
in time.  
This cannot be handled by the minimum-mass model nor by steady-state
models. Observations of circumstellar disks
surrounding T Tauri stars support the view of disks having a limited
life-span and characterized by continuous changes during their
life. This evidence has led to a large consensus about the nebular
origin of the Solar System. 
In addition, one also needs to describe the global evolution of
the solid material which constitutes, together with the gas, the
protoplanetary disk. 
The knowledge of this distribution and
its time evolution is important both to understand how planets form and to calculate the migration rate.

The time evolution of the surface density of the gas
$\Sigma$ is given by the familiar equation (e.g., Stepinski \& Valageas 1997):
\be
\frac{\partial \Sigma}{\partial t} -\frac{3}{r}\frac{\partial
}{\partial r}\left[ r^{1/2}\frac{\partial }{\partial r}\left(
r^{1/2}\nu _{t}\Sigma \right) \right] =0
\label{gasevol}
\ee
where $\nu_{\rm t}$ is the turbulent viscosity. 
Since $\nu_{\rm t}$ is
not an explicit function of time, but instead depends only on the
local disk quantities, it can be expressed as $\nu_{\rm t}=\nu_{\rm
t}(\Sigma,r)$ and Eq.(\ref{gasevol}) can be solved subject to
boundary conditions on the inner and outer edges of the disk. The
opacity law needed to compute $\nu_{\rm t}$ is obtained from
Ruden \& Pollack (1991). Then, Eq.(\ref{gasevol}) is solved by means of an
implicit scheme. Note that the evolution of the gas is computed
independently of the evolution of particles (which only make $\sim
1\%$ of the gas mass). Next, from $\Sigma(r,t)$ we can algebraically
find all other gas disk variables.

Next, as described in Stepinski \& Valageas (1997) the evolution of the surface density
of solid particles $\Sigma_{\rm s}$ is given by:
\be
\frac{\partial \Sigma_{\rm s} }{\partial t} = \frac{3}{r} \frac{\partial
}{\partial r} \left[ r^{1/2} \frac{\partial}{\partial r} (\nu_{\rm s}
\Sigma_{\rm s} r^{1/2}) \right] + \frac{1}{r}\frac{\partial }{\partial
r} \left[ \frac{2r \Sigma _{s} \langle \overline{v}_{\phi}
\rangle_{s}} {\Omega _{k}t_{s}} \right] .
\label{solidevol}
\ee
where the
effective viscosity $\nu_{\rm s}$ is given by:
\be
\nu_{\rm s} = \frac{\nu_{\rm t}}{{\rm Sc}} \hspace{0.4cm} \mbox{with}
\hspace{0.4cm} {\rm Sc} = \left( 1+\Omega _{\rm k} t_{\rm s} \right)
\sqrt{1+\frac{\overline{\bf v}^2}{V_{\rm t}^2}} .
\label{Schmidt1}
\ee
where Sc is the Schmidt number, which measures the coupling
of the dust to the gas turbulence,
${\bf v}$ is the relative
velocity between a particle and the gas, $V_{\rm t}$ the turbulent
velocity, $\Omega _{\rm k}$ is the Keplerian angular velocity 
and $t_{\rm s}$ the so-called stopping time. 

The first diffusive term in Eq. (2) is similar to Eq.(\ref{gasevol}), 
the second advective term in Eq.(\ref{solidevol}) arises from the lack
of pressure support for the dust disk compared to the gas
disk. Thus, it is proportional to the azimuthal velocity difference
$\overline{v}_{\phi}$ between the dust and the gas. The average
$\langle .. \rangle_{s}$ refers to the vertical averaging over the
disk height weighted by the solid density. We refer the reader to
Stepinski \& Valageas (1997) for a more detailed presentation, see also Kornet et al.(2001).

In addition to the radial motion described by Eq.(\ref{solidevol}),
the dust surface density also evolves through evaporation/condensation
and coagulation. 
The method used to incorporate these processes into 
calculations relies on keeping Eq. (\ref{solidevol}) as the principal 
mathematical description of the global evolution of particles, 
but freeing its parameters from constraints of single-size and 
thermal indestructibility assumptions.
Thus, the method can be characterized as solving a radial
advection-diffusion problem modulated by coagulation, with
the possibility of a cut-off by the evaporation. This method requires
that the mass distribution of particles at any given radial
location of a disk is narrowly peaked about a mean value particular
for this location and a given time instant. Such an assumption
may appear quite arbitrary; however, it has a reasonable physical
justification as collisions with small particles do not significantly
increase the size of a test particle and the bulk of the
solid mass is concentrated in the largest particles. This is supported by numerical simulations (Mizuno et al. 1988) of grain growth in protoplanetary disks which clearly
show that although a broad size distribution is maintained, most
of the mass is nevertheless always concentrated in the largest
particles.
Therefore,
we assume that the mass distribution of particles at any given radial
location r is narrowly peaked about the mean value 
$m_{\rm d}(r,t)=(4/3) \pi s^3(r,t) \rho_{\rm s} $, 
where s(r,t) is the size. The goal is to evaluate
the functional dependence of s on r and t. Particles are assumed
to be spheres with a bulk density of  $\rho_{\rm s} $ . 
The density of matter concentrated into solid particles is $\rho_s(r)$. 
Then, within this approximation,
coagulation does not influence the dust surface density $\Sigma_{\rm
s}$ since it conserves the total mass of solids. Thus, the coagulation
of solid particles only appears through the evolution of the radial
distribution $s(r,t)$ of the typical size of the dust grains. We
model this process as in Stepinski \& Valageas (1997). We only
consider collisional coagulation and we disregard gravitational
interactions which would come into play at late times when large
planetesimals have formed.

On the other hand, we take into account the evaporation of solid
particles which takes place at the radius $r_{\rm evap}$ where the
temperature reaches $T_{\rm evap}$. We also include the condensation
of the vapor below $T_{\rm evap}$ onto the solid grains. 
%
%
As In DP3, we consider only one species of solid particles:
high-temperature silicates with $T_{\rm evap}= 1350$ K and a bulk
density $\rho_{\rm bulk} = 3.3$ g cm$^{-3}$. Thus, in our simplified
model we follow the evolution of three distinct fluids: the gas, the
vapor of silicates and the solid particles.

In this fashion, we obtain the radial distribution of the planetesimal
swarm after $10^7$ yr. This yields the surface density of solids
$\Sigma_{\rm s}(r,t)$ and the mid-plane solid density $\rho_{\rm
s}(r,t)$. We also obtain the size distribution $s(r,t)$ reached by
hierarchical coagulation. Of course, at these late times where
planetesimals have typically reached a size of a few km or larger,
gravitational interactions should play a dominant role with respect to
coagulation. However, if these interactions do not significantly
affect the radial distribution of solids (note that the radial
velocity of such large particles due to the interaction with the gas
is negligible) we can still use the outcome of the fluid approach
described above to study the migration of giant planets, as detailed
below.

As in Stepinski \& Valageas (1997), we 
consider an initial gas surface density of the form:
\be
\Sigma_0(r)= \Sigma_1 \left[1+(r/r_1)^2\right]^{-3.78} + \Sigma_2
(r/1{\rm AU})^{-1.5} .
\label{Sigma0}
\ee
The quantities $\Sigma_1$, $r_1$ and $\Sigma_2$ are free parameters
which we vary in order to study different disk masses. 
The values we
use are given in Table 1, where $M_{\rm d}$ is the gas disk mass (in units of
$M_{\odot}$), $J_{50}$ is the disk angular momentum (in units of
$10^{50}$ g cm$^2$s$^{-1}$), $\Sigma_1$ and $\Sigma_2$ are in g
cm$^{-2}$ and $r_1$ is in AU. The first term in Eq.(\ref{Sigma0})
ensures that there is some mass up to large distances from the star,
while the second term corresponds to the central concentration of the
mass and sets the location of the evaporation radius. Note however
that the evaporation radius for the high-temperature
silicates we study here remains of the order of $0.1$ AU. As previously explained, 
we consider only one species of solid: high-temperature
silicates with $T_{\rm evap}=1350$ K. We initialize the dust subdisk
at time $t=10^4$ yr (i.e. after the gas distribution has relaxed
towards a quasi-stationary state) by setting the solid surface density
$\Sigma_{\rm s}$ as: $\Sigma_{\rm s}(r,t=10^4 yr)=6\times 10^{-3}\Sigma(r,t=10^4 yr)$, 
similarly to Stepinski \& Valageas (1996,1997), in order to account for cosmic abundance. 

\begin{table} 
\begin{center} 
\caption{Properties of the initial gas disk}
\begin{tabular}{|l|l|l|l|l|} \hline
$M_{\rm d}$      &  $J_{50}$  &  $\Sigma_1$ &  $r_1$  &  $\Sigma_2$ \\ \hline
$10^{-1}$  &  $911$     &  $22$       &  $200$  &  $100$      \\ \hline
$10^{-2}$  &  $85$      &  $1.7$      &  $200$  &  $100$      \\ \hline
$10^{-3}$  &  $5.5$     &  $1.2$      &  $50$   &  $30$       \\ \hline
$10^{-4}$  &  $0.46$    &  $0.2$      &  $50$   &  $2.8$      \\ \hline
\end{tabular}
\end{center} 
\label{table1} 
\end{table}

The assumed initial surface density profile is
arbitrary. Fortunately, the specific form of the profile does
not influence the subsequent evolution of the gas, 
inasmuch as the process governed by Eq. (\ref{gasevol}) is diffusive
in nature and the details of initial distribution are 
forgotten after a time short in comparison with the evolutionary
timescale. 

Taking advantage of the ``memory" of the
gaseous component, we introduce solid particles into the
calculation only after $10^4$ yr during which
the gas evolves alone to remove the arbitrariness in the
initial conditions. Upon introduction, the solid particles all
have the same size, $s = 10^{-3} cm$. 

The calculations are carried out for up to $t =10^7$
yr, a period of time equal, within an order of magnitude, to
observationally deduced lifetimes of protoplanetary disks
(Strom \& Edwards 1993).

As shown in DP3, after times of the order of $10^6-10^7 {\rm yr}$,
the coagulation process gives rise to large particles ($>10^5 {\rm
cm}$) which have a small radial velocity, and hence a negligible radial
motion.  This leads to a freezing of the solid surface density to the
value reached at times of the order of $10^7 {\rm yr}$.  In other
words, once solids are in the form of planetesimals, the gas coupling
becomes unimportant and the radial distribution of solids does not
change
and can only change
on a much longer timescale by processes not considered in our model
(like mutual gravitational interactions between planetesimals first proposed
 by \citeNP{Safronov1}).
\footnote{Note that $10^6$ yr is the time required for $\Sigma_{\rm
s}$ (or $\rho_{\rm s}$) to converge everywhere. However, this
convergence is not uniform and can be achieved on a time scale as
short as $10^4$ yr in the innermost disk where the dust density is
highest.} This is why we do not need to calculate its evolution
for times longer than $10^7$ years.



\subsection{Migration model}

With regard to the migration model, 
we consider a planet revolving around a star of mass $M_{\ast}=1
M_{\odot}$. The equation of motion of the planet can be written as
\be
{\bf \ddot r}= {\bf F}_{\odot} + {\bf R}
\ee
where the term ${\bf F}_{\odot}$ represents the force per unit mass
from the star, while ${\bf R}$ is the dissipative force (i.e. the
dynamical friction term--see \citeNP{Melita1}). 
Then, the
frictional drag on the test particles may be written as
\be
{\bf R}=-k_{\parallel}v_{1 \parallel} {\bf e_{\parallel}}-
k_{\perp}v_{1 \perp}{\bf e_{\perp}} \label{eq:dyn}
\ee
where ${\bf e_{\parallel}}$ and ${\bf e_{\perp}}$ are two unit
vectors parallel and perpendicular to the disk plane, and
$k_{\parallel}v_{1 \parallel}$ and $k_{\perp}v_{1 \perp}$
are given in DP3.
We assume a
disk-shaped matter distribution with constant velocity dispersions
$\sigma_{\parallel}$ (parallel to the plane) and $\sigma_{\perp}$
(perpendicular to the plane) and with a ratio simply taken to be 2:1
(i.e. $\sigma_{\parallel}$=$2 \sigma_{\perp}$).
Since the damping of eccentricity and inclination is more rapid than
radial migration (\citeNP{Ida1}; \citeNP{Ida2}; DP1), we only deal
with radial migration and we assume that the planet has negligible
inclination and eccentricity ($i_{\rm p} \sim e_{\rm p} \sim 0$) and
that the initial distance from the star to the planet is $5.2$ ${\rm
AU}$. 
The planet was set at an initial distance of 5.2 ${\rm
AU}$ for similarity with choices made in previous papers (e.g., Murray et al. 1998, Trilling et al. 1998)

In the calculation, we consider a two-component system, consisting of one 
protoplanet and many equal-mass planetesimals. 
The velocity dispersion of planetesimals in the neighborhood of the
protoplanet depends on the mass of the protoplanet. When the mass of the
planet, $M_{\rm p}$, is $\le 10^{25}$ g, the value of $<e^2_{\rm m}>^{1/2}$
( $e_{\rm m}$ being the eccentricity of the planetesimals) is
independent of $M_{\rm p}$ therefore:
\begin{equation}
e_{\rm m} \simeq 20 (2 m /3 M_{\odot})^{1/3}
\end{equation}
(Ida \& Makino 1993) where $m$ is the mass of the planetesimals. When 
the mass of the planet reaches values larger than $10^{25}$-$10^{26}$ g at
1 AU, $<e^2_{\rm m}>^{1/2}$ is proportional to $M_{\rm p}^{1/3}$:
\begin{equation}
e_{\rm m} \simeq 6 (M_{\rm p}/3M_{\odot})^{1/3}
\end{equation}
(Ida \& Makino 1993).
As a consequence, the dispersion velocity in the disc is also
characterized by two regimes, being connected to the eccentricity
by the equation:
\begin{equation}
\sigma_{\rm m} \simeq (e_{\rm m}^2+i_{\rm m}^2)^{1/2} v_{\rm c}
\end{equation}
where $i_{\rm m}$ is the inclination of planetesimals and
$v_{\rm c}$ is the Keplerian circular velocity.
Following Stern (1996) and Del Popolo et al. (1999) we assume that $\langle i_{\rm m}^2 \rangle = \langle
e_{\rm m}^2 \rangle/4$. 
In the simulation we assume that the planetesimals all have  
equal masses, $m$, and that $m<< M_{\rm p}$.
This assumption does not affect the results, since dynamical friction does not depend
on the individual masses of these particles but on their overall 
density (see DP3).
Therefore, we do
not need to follow the evolution of the size distribution of
planetesimals. We merely use the planetesimal density $\rho_{\rm s}$
reached after $10^7$ yr, assuming that the height of the planetesimal
disk does not evolve significantly.

Other details, and a discussion about the back reaction of the planet on the swarm, 
can be found in DP1, DP2, DP3.

\subsection{Further assumptions and the distribution calculation method}

In order to determine the distribution of semimajor axes and mass, 
it is necessary to investigate the dependence of the giant planet migration
on the properties of the protoplanetary disk and planet mass $M_{\rm p}$.
Similarly to what we did in DP3, we integrated the model
introduced in the previous sections for several values of the initial
disk surface density (i.e. several disk masses), 
different values of $\alpha$, and planet mass. In this way, it was possible to obtain 
a semimajor axis corresponding to a given value of $\alpha$, disk mass, and $M_{\rm p}$. 
As previously reported, the assumptions and model are the same as in DP3.

Since, similarly to DP3,  we are mainly interested in studying the planet migration due to the interaction
with planetesimals, we assume that the gas is almost dissipated when the planet starts its migration.
We know that, usually, 
while the gas tends to be dissipated,
(some evidence shows that the disk lifetimes range from $10^5$ yr
to $10^7$ yr, see Strom et al 1993; Ruden \& Pollack 1991), the coagulation
process induces an increase of the density of solid particles with
time  and gives rise to objects of increasing dimensions.
%
%
Clearly the effect of the presence of gas should be that of
accelerating the loss of angular momentum of the planet and to reduce
the migration time. In our case, there is still gas after $10^7 {\rm yrs}$, but it is in  
quantity inferior to that of planetesimals, especially in the case of lower mass discs, in 
which it can be even two order of magnitude less than planetesimals. Moreover, as noticed by 
Kominami \& Ida (2002) dynamical friction and gravitational gas drag are essentially the 
same dynamical process and the effect of the gravitational gas drag is incorporated in 
that of dynamical friction. In our case, other forces, such as aerodynamical gas drag, can be neglected when 
compared to gravitational gas drag (Kominami \& Ida 2002).

\begin{figure}
\centerline{\hbox{
\psfig{figure=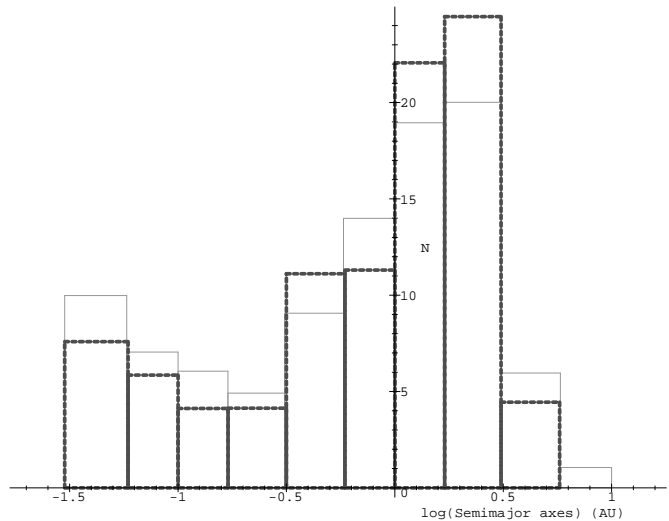,width=12cm,angle=270}
}}
\caption[]{Orbital semi-major axes for observed (solid line), and model (dashed thick line) Extra-solar Giant Planets (EGPs). 
The solid line represents the semi-major axes distribution obtained with the data given at
www.exoplanets.org, while the dashed thick line is the distribution obtained with the model of this paper. 
}
\end{figure}

\begin{figure}
\centerline{\hbox{
\psfig{figure=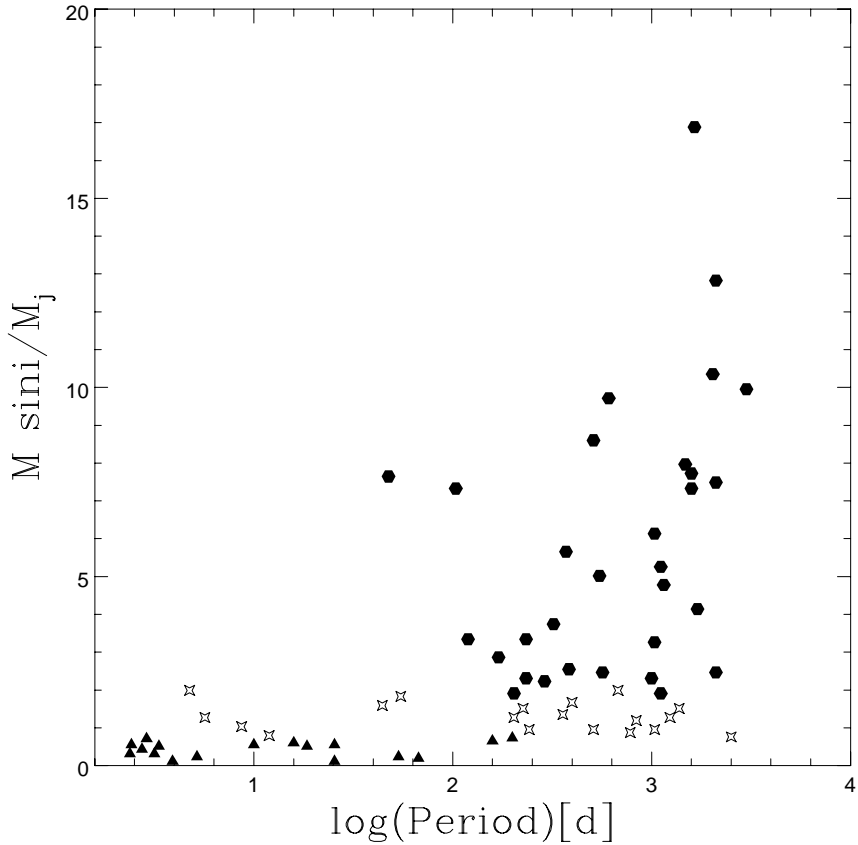,width=12cm,height=12cm}
}}
\caption[]{Mass versus period of planets calculated according to the model of the current paper. 
Filled symbols represent massive ($M_{\rm p} \geq 2 M_{\rm J}$) planets, open symbols 
represent intermediate-mass planets ($0.75 \leq M_{\rm p} \leq 2 M_{\rm J}$), and 
filled triangles lighter ($M_{\rm p} \leq 0.75 M_{\rm J}$) planets
}
\label{Figperiod1}
\end{figure}

\begin{figure}
\centerline{\hbox{
\psfig{figure=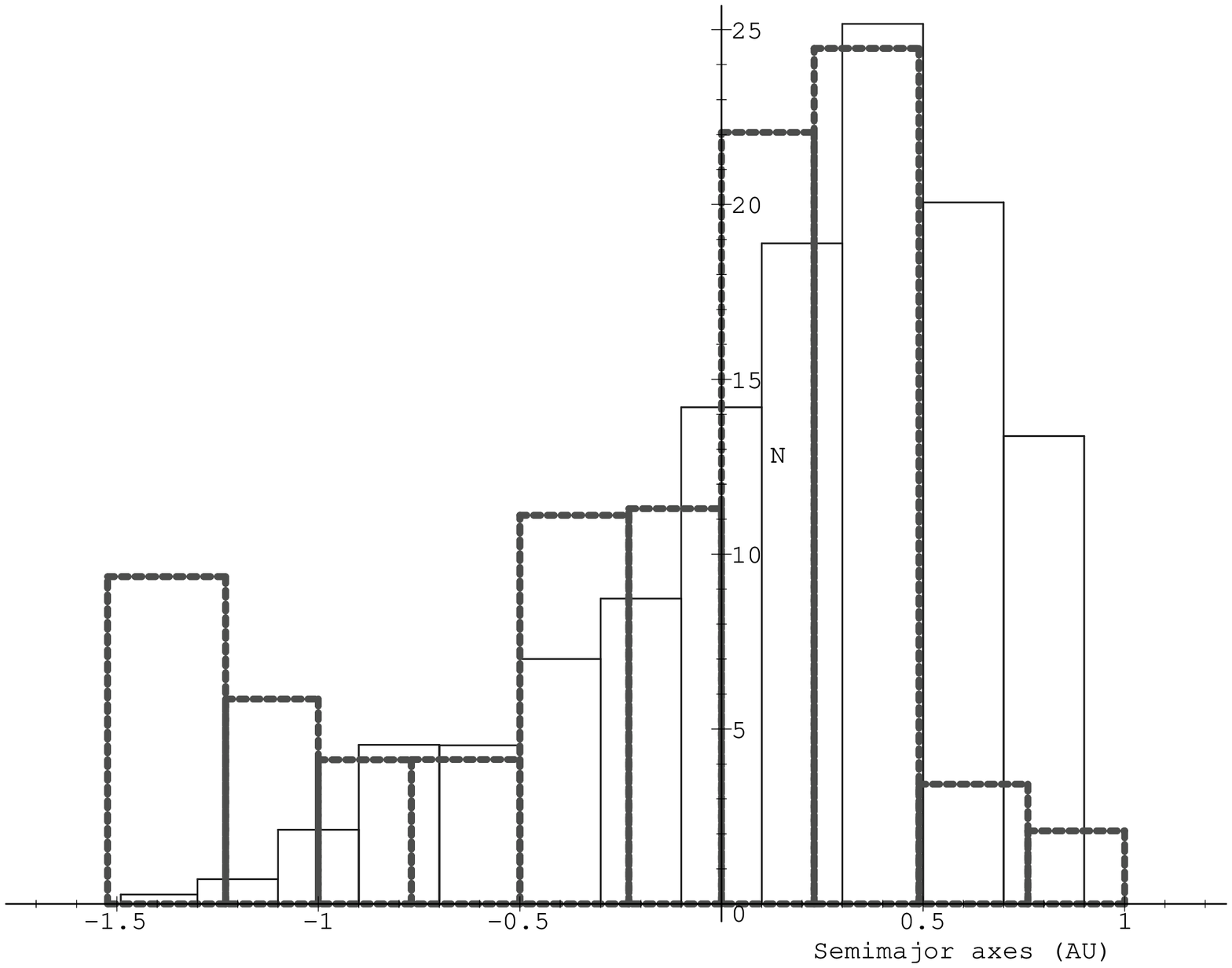,width=12cm,angle=270}
}}
\caption[]{Orbital semi-major axes for EGPs obtained in the model of this paper (thick dashed line) compared to the 
model of Trilling et al. (2002) (solid line) . 
}
\end{figure}

In order to calculate the fraction of planets in terms of semi-major axes and mass, 
we used the model described in the previous section,
and we have performed a statistical study of planet migration in systems in which both disk
mass and disk viscosity 
are varied in turn. 
The initial orbital semi-major axes of all planets are 5.2 AU. 
Initial planet masses are taken to be fractional values of Jupiter masses, 
between 0.1 and 18 Jupiter masses (with intervals of $0.1 M_{\rm J}$), considering one planet per system.  
As in Trilling et al. (2002), for the varied parameters, we
adopt nominal values and vary each parameter over a given range.
In the case of disk mass, we adopt a nominal disk mass of $0.02 M_{\odot}$ (after Beckwith
et al. 2000), and calculate model cases between $10^{-4} < M_{\rm d} < 0.6 \times 10^{-1} M_{\odot}$ (with intervals of $0.01 M_{\odot}$).
This range spans disk masses from less than the minimum mass disk for a system
containing a Jupiter, to a disk so massive that direct binary star production is more likely
than giant planet formation (see, for example, reviews by Beckwith \& Sargent 1993 and
Beckwith et al. 2000). 
Similarly, we use a nominal $\alpha$ viscosity of $3 \times 10^{-3}$ (Shakura \& Sunyaev
1973, Trilling et al. 1998) and test parameter values 
in the range of $10^{-4} < \alpha < 10^{-1}$ (with intervals of $0.01 \alpha$)
(the values used by Trilling et al. 2002 are 
almost the same, $10^{-4} < \alpha < 9 \times 10^{-2}$). 
Each model run has one parameter varied while the other two are given their nominal values. 
This very fine sampling of disk parameters allows us to simulate a continuum of physical properties of planet-forming
systems. 

For each value of the planet mass, we obtain a three-dimensional grid of models, based on the 
nominal values and spanning the range of parameter space.
To produce a Gaussian statistical population 
the significance of each model run was weighted by
the Gaussian probability of the value of the varied parameter according to the standard
Gaussian formalism:
\begin{equation}
P(y) =\frac{1}{\sqrt{2 \pi} \sigma} e^{-\frac{1}{2}\left[(y-y_0)\right]^2/\sigma^2}
\end{equation}
where $y=\log{x}$ and $y_0 = \log{x_0}$, where $x$ is equal to the value of the varied parameter,
$x_0$ is the nominal value for the parameter being varied, and where $\sigma$ (describing
the width in log space of the Gaussian parameter distribution) is equal to 0.5. In
other words, the half-width of the Gaussian physical parameter distribution is a factor
of $\sqrt{10}$. 
\footnote{Thus, a parameter value which is $\sqrt{10}$ times greater or less than
the nominal value has a probability one half of the probability that of the nominal value:
$P(y_k) = \frac{1}{2}P(y_0)$ for $y_k = y_0 \pm 0.5$ 
}
We use this Gaussian weighting in all results discussed in this paper. Weighting is
achieved by assigning a significance to each 
planet as $P(y)$.
Thus, planets from extreme disks which survive but whose initial conditions are extremely
unlikely are not overrepresented in the statistical results. Note that because probabilities
are assigned after all model runs are completed and are simple functions of the nominal
values and a description of a Gaussian distribution, other probability weightings can be
installed to represent other experiments. Future work includes deriving both $x_0$ and $\sigma$
more directly from observations (see, e.g., Gullbring et al. 1998; Hartmann et al. 1998)
and testing non-Gaussian probability distributions.


%
%
%
%
%
%
%
%
%

\section{Results}

\subsection{Semi-major axis distribution}

The results of the model of the previous section are plotted in Figs. 1-3.
In the first, we represent the distribution of semi-major axis 
of planets.
In Fig. 1, the light solid line represents the semi-major axis distribution obtained with the data given at
www.exoplanets.org (see also Fig. 3 of Marcy et al. 2003)
while the dashed thick line is the distribution obtained with the model of this paper. 
%
%


The plot in Fig. 1 shows that we obtain a reasonable agreement with the data. 
However, the main point of Fig. 1 is to show that the observed
distribution of orbital periods can be easily recovered from our
model, and with reasonable values for $\alpha$
and $M_{\rm d}$.

The plot shows that at $a \simeq 0.05$ AU there is a pile-up of planets. This pile up  
seems not to be an
artifact of observational selection (see \citeNP{Kuchner1} and subsection 3.3),
and it suggests that some mechanism must be responsible for halting the migration 
of planets at an orbital period of 3 days.  

As stressed by Kuchner \& Lecar (2002), phenomena like the interaction between
two planets and a star, that can leave a planet trapped by stellar
tides into a circular orbit of $\simeq 0.04$ AU (\citeNP{Rasio1}), or
halting of planet migration by loss of mass to the star
(\citeNP{Trilling1}), are rare. The migration through resonant
interaction with planetesimals in the disk (\citeNP{Murray1})
provides a natural way of halting the planet but unfortunately to
change the orbit of a Jupiter-mass planet requires approximatively a
Jupiter mass of planetesimals and a too massive disk. The other
possibility to explain the pile-up of planets seen in Fig. 1 (see also DP3) is a gas disk truncated at a
temperature of 1500 ${\rm K}$ by the onset of Magneto-Rotational
instability (\citeNP{Kuchner1}).  Thus, disk temperature
determines the orbital radii of the innermost surviving planets.  In
our model, the distribution of internal planets is naturally connected
to the evaporation of planetesimals at very short radii. As stressed
in DP1, DP2, this model does not have the drawback of Murray (2002), namely
that of requiring a too-large disk mass for migration, and at the same
time it has the advantage of Murray (2002) of having an intrinsic
natural mechanism that provides halting of migration.
As shown in DP3, the same model used in this paper is able to explain the peak at 3-4 days 
for the orbital period.

A second feature visible in Fig. 1
is a minimum near 0.3 AU, indicating a paucity of planets at that distance. 
One interpretation of this minimum is that the migration mechanism rarely 
allows them to stop near 0.3 AU (Jones et al. 2003). 
There is the suggestion of a rise
for semi-major axes larger than 0.3 AU, out to 3 AU. Beyond 3 AU, 
the Doppler surveys have had inadequate lifetimes ($ \simeq 8$ yr) to securely detect 
planets in the correspondingly long orbital periods (the decrease observed after 3 AU is due at least in part to observational bias). 
\footnote{Indeed, there is poor detectability of planets for semi-major axes greater than 2 AU, where planets with 
$M_{\rm p} < 1 M_{\rm J}$ are currently difficult to detect, given the precision and duration of surveys.} 
Thus, observations summarized in Fig. 1 strongly indicate the existence of a rise in the 
distribution of planets with increasing orbital distance from the host star. 
The increasing numbers of planets in larger orbits is consistent with models 
that invoke orbital migration, like that of this paper, and a contemporaneous clearing of the gaseous disk 
to explain the final positions of giant planets (Trilling et al. 2002). 
If we have planets that initially reside at 5 AU or at larger distances, 
and if the disk clears its gas 
on a time scale shorter than the migrational time scale, 
then giant planets beyond 5 AU could be more numerous than the extrasolar planets discovered so far. 

%
%

In conclusion, we can tell that the disk model and migration mechanism used 
in this paper
are able not only to predict the planet distribution at short semi-major axes,
with a peak at 3-4 days, but also the distribution of planets at larger distances.

\subsection{Mass versus period}

In Fig. 2, we plot planet mass vs. orbital semi-major axis. There is a lack of massive planets 
having small semi-major axes. Within $\simeq 0.3$ AU ($\simeq 60$ days) there are no planets with masses larger than 
$\simeq 4 M_{\rm J}$ and moreover there are few massive planets with semi-major axis $< 1$ AU. 

With now more than 100 extra-solar planets discovered,
the lack of short-period massive
planetary companions becomes even clearer, as seen
in Fig. 2. When we neglect the multiple-star systems
\footnote{In these systems the planetary formation or
evolution could follow different paths (Zucker \& Mazeh
2002; 
Udry et al. 2003)
}, 
a complete void of candidates is observed in the diagram for masses larger
than $\simeq 2 M_{\rm J}$ and periods smaller than $\simeq 100$ days. 

This feature can be explained by several processes:\\
1) Type II migration.\\
2) Processes related to planet-star interactions 
when the planet reaches the central regions (see Udry et al. 2003).\\
3) Multi-planet chaotic interactions (Rasio \& Ford 1996;
Weidenschilling \& Marzari 1996).

In relation to type II migration, the feature can be explained assuming 
that the migration process has a low efficiency (planets rarely migrate inward) or too high an efficiency (planets 
migrate all the way into the star) for massive planets (mass $>4 M_{\rm J}$). The first alternative (low efficiency) 
is the most probable. This view is supported by simulations of single-planet migration and by the model of this paper.
More massive planets create larger gaps than smaller planets 
(Trilling et al. 1998). For the most massive planets, the length
of time it takes to fully form the gap can be quite long and 
during gap formation there is no Type I migration since the planet has already opened a gap; there is
also little Type II migration because most Lindblad resonances already fall within the
growing gap. 
If the timescale for gap growth is longer than the
disk lifetime, then the planet will not migrate far. Therefore, overall, close-in
surviving model planets have smaller masses and distant surviving models planets have
larger masses.

The explanation of the lack of short-period massive
planetary companions is similarly explained in our model.  
As discussed in DP2, DP3 (in connection with the back reaction of the planet on the swarm),
when the protoplanet becomes massive enough ($10^{25}-10^{26}$ g) to influence the velocity distribution of small planetesimals, the random velocities and velocity dispersion of planetesimals are heated by the protoplanet and become larger than in the early stage.   
Furthermore, the protoplanet would scatter neighboring planetesimals and give rise to a gap in the spatial distribution of planetesimals (see Fig. 3 of Ida \& Makino 1993, and Fig. 1a of Tanaka \& Ida 1997). 
The effect on the drift velocities can be easily predicted observing that 
the increase in velocity dispersion of planetesimals around the protoplanet decreases the
dynamical friction force and consequently 
increases the migration time-scale. 

As shown in DP3, the drift
velocity behaves as:
\begin{equation}
\frac{d r}{dt} \propto \left\{
\begin{array}{lc}
M & M \le 10^{25} g  \\
constant & M > 10^{25} g
\end{array}
\right.
\end{equation}
The linear dependence on the planet's mass of the drift velocity corresponds to the $type$ I drift in the density wave approach (Ward 1997), while the part independent of the planet's mass corresponds to $type$ II drift. 
The transition between the two regimes entails a velocity drop of between one to three orders of 
magnitude (according to the value of $\alpha$).
As the threshold is exceeded,
the motion fairly abruptly converts to a slower mode in which the
drift velocity is independent of mass
\footnote{The conclusion that migration timescales are independent of planet
mass is valid in the limit that the planet's mass is small compared to the disk
mass. 
In the
case of a finite mass ratio between planet and disk, however, there is a feedback torque from the
planet onto the disk, and the planetary migration timescale is seen to be a function of planetary
masses (Lin \& Papaloizou 1986; Trilling et al. 1998; Bryden et al. 2000; Nelson et al. 2000;
Kley 2000; D'Angelo et al. 2002). 
In the regime in which we are working, planetary migration
timescales are functions of planet masses in the sense that, in general, smaller planets migrate
faster and more massive planets migrate more slowly. 
}. 
%
%
The phenomenon is equivalent to that predicted in the density wave 
approach (Goldreich \& Tremaine 1980; Ward 1997). This leads to similar conclusions obtained by Trilling et al. (2002), with a type II migration model: higher mass planets migrate on longer timescales. More massive 
planets preferentially form in the outer regions where a larger amount of building material is available, and that from those 
distances they need longer times to migrate towards the inner region. 

we have to add some hints on the way migration was calculated 
in the case the disc mass, $M_{\rm d}$, is smaller than that of the protoplanet. 
One can think that less massive discs are not able to move a more massive planet. This idea 
is not completely correct.
As we know, when the planet mass is less than or comparable to the local disc mass,  
the planet behaves as a representative particle in the disc, as shown by Lin \& Papaloizou (1986).
When the protoplanet has a mass larger than that of the disc, the satellite acts as a dam against the 
viscous evolution of the disc, and can lead to a substantial change in the disc structure in the 
vicinity of the planet. The coupled disc-planet evolution in this case has been studied by Syer \& Clarke (1995) and 
by Ivanov et al. (1999), who showed that the inertia of the planet plays an important role in this case (see Nelson et al. 2000, Eq. 9). The effect of the interaction between the secondary and the disc leads to accumulation of the disc matter 
in the region behind the protoplanet (see Eq. 12 and 39 of Syer \& Clarke 1995). In order to calculate the migration 
in our model, we used the surface density given in Ivanov et al. (1999).

As shown in Fig. 2, 
another very interesting feature comes out of the period/separation 
distribution: there is a shortage of
planets with periods between roughly 10 and 100 days.
This observational property is mainly due to the light
candidates ($M_{\rm p} \sin i \leq 2M_{J}$). As seen in the previous section,
massive planets orbiting single stars are almost exclusively
found on longer-period orbits. They probably form
and stay further out. On the
other hand, lighter planets ($ 0.75 \leq M_{\rm p} \sin i \leq 2M_{J}$), are found at all distances from
their stars, unlike massive planets. 

%
%


A third 
feature emerging from the mass-period
diagram is the apparent lack of very light planets
($M_{\rm p} \sin i \leq 0.75 M_{\rm J}$) with periods larger than $\simeq$ 
100 days.
Assuming that it is not produced by
small number statistics,
an explanation of this feature has been advanced by 
Masset \& Papaloizou (2003), Udry et al. (2003).
Masset \& Papaloizou (2003) studied the effect
of co-orbital corotation torque on migrating protoplanets.
In particular they showed that if the
mass deficit created by the radial drift with the planet of
the material trapped in the co-orbital region is larger than
the planet mass, the migration rate undergoes a runaway
which can rapidly vary the protoplanet semi-major axis
by a large amount (50\% over a few tens of orbits). 
This typically corresponds to planet masses in the
sub-Saturnian to Jovian mass range embedded in massive
protoplanetary disks.

Moreover, the simulations by Masset \& Papaloizou
(2003) show that the limit in planetary mass for the
appearance of the runaway regime is very steep (their
Figs. 12-14). 
There is a critical mass
relatively well-constrained ($M_{\rm crit} \simeq 1 M_{\rm J}$) under which
runaway is likely provided that the protoplanetary disk
is not too lightweight, and above which runaway is impossible.
%
%
In Fig. 2, it is shown that in the calculations of this paper, 
light planets, $M_{\rm p} \sin i \leq 0.75 M_{\rm J}$, are found
predominantly at periods shorter than 100 days, but some planets 
are found also at larger distances.
In others words, following the mass trend seen for other planets, 
smaller mass planets migrate to smaller periods, but differently from
observations, some light planets, $M_{\rm p} \sin i \leq 0.75 M_{\rm J}$, 
are observed also at periods larger than 100 days. 
This suggests that if there is a larger probability of finding light planets 
with smaller periods, it is also possible that some light planets can 
be found on larger orbits, depending on particular characteristics of 
the disk. The fact that in the sample of $\simeq 100$ planets we have now
the light planets are found exclusively on periods shorter than 100 days can be
related to the fact that the detection probability decreases with increasing distances and 
with decreasing mass: light planets $M_{\rm p} \sin i \leq 0.75 M_{\rm J}$ are more and more 
difficult to detect at larger distances.  
In other words, this feature could be related to the observational
bias (see subsection 3.3) inherent to the radial-velocity technique for
planet search that makes the detection more difficult for
distant and/or lighter planets (but see also Udry et al. 2003). 

P.J. Armitage (private communication)
suggested that the probability that  
%
the observational evidence of the
lack of planets with $M_{\rm p} \sin i \leq 0.75 M_{\rm J}$ with $P > 100$ days
is a real effect is only marginal. 
This comes from the fact that plotting the mass vs semi-major axis
distribution, the lower edge of the
distribution nearly follows
the line defined
by constant stellar radial velocity amplitude (i.e. 
$ \propto \sqrt{a}$). If there is a lack of low mass planets present
at $~1$ AU separation, there are only maybe one or two
planets ``missing", leading to the conclusion that the planet lack quoted is definitely connected to small number statistics.

\subsection{Bias and significance of the data.}

Here we add some statements about the significance of the observational data, in general.
No massive planets ($M \sin{i} \geq 2 M_{\rm J}$) are found on short periods orbits ($P \leq 100$ d) around single stars. 
This feature has been pointed out by Udry et al (2002) and Zucker \& Mazeh (2002).
The same authors verified its statistical significance and examined the possible influence
of binarity on the mass-period relation of exoplanets, an indication
of potentially different formation and evolution processes for planets in binaries and planets around single stars.
The observed feature is most probably not an observational bias since these objects are the easiest to detect. 

The second important feature that emerges from observations is a planet shortage in
in the 10 and 100-day periods, the ``period valley", following Udry et al. (2003). 
The features appearing in the planetary period distribution peak at short periods and rising at 
intermediate periods seem significant and not due to observational biases. 
The peak at short periods is formed by the pile-up of migrating
planets, stopped close to the central star. The rise of the distribution
at longer periods is more clearly emerging thanks to the increase of the
timebase of the radial-velocity surveys. 
More and more long period planets are being detected. This rise is significant in the
sense that, for a given mass range, similar planets on shorter period
orbits would have been easier to detect than the actual
ones. The other, previously quoted feature is the lack of 
planets with $M_{\rm p} \sin i \leq 0.75 M_{\rm J}$ with $P > 100$ days. This feature could
be related to the observational bias inherent to the radial-velocity technique for planet search
that makes the detection more difficult for distant and/or lighter
planets. In order to check the statistical significance of the effect, Udry et al. (2002)
performed Monte-Carlo simulations. They conclude that the considered region is indeed empty 
of planets with a confidence level of about 99.97 \%. 
This conclusion is only valid if the observational biases are
correctly taken into account. 
For example, activity-induced radial-velocity jitter may
screen planet detection. 
Another concern relates to the difficulty
encountered when trying to actually derive the values for the
orbital elements, which are different to just detecting radial-velocity
variability. 

As previously observed, a different point of view is that 
the probability that the effect is a real effect is only marginal. 
This comes from the fact that plotting the mass vs semi-major axis
distribution, the lower edge of the
distribution is close to following the line defined
by constant stellar radial velocity amplitude (i.e. 
$ \propto \sqrt{a}$). 

More quantitative constraints for the different mass
regimes will require a better detection by the surveys,
to diminish the effect of the observational bias in the distributions
and thus increase the confidence we can put in the derived
trends, especially for the lower-mass planets.




\subsection{Comparison with other studies}

A comparison with Trilling et al. (2002) is performed in Fig. 3. In this plot 
the thick dashed line represents the model of this paper while the light solid line Trilling's prediction.
As previously reported, using a model assuming a simple impulse approximation
for the type II migration, no ad-hoc stopping mechanism
in the center and neglecting type I migration,
Trilling et al. (2002) show that, for identical initial disk
conditions, close-in surviving pseudo-planets have smaller
masses and distant surviving pseudo-planets have larger
masses. They also find that planets forming farther out
migrate less rapidly and that higher-mass planets migrate
on longer timescales. Thus, there should be more massive
planets at intermediate and large semi-major axes than
found close-in.
Conversely, the population of short-period planets should be dominated by smaller-mass planets,
as is indeed observed.

Similarly to Trilling et al (2002), the present paper concludes that 
close-in surviving pseudo-planets have smaller
masses and distant surviving pseudo-planets have larger
masses. 
Differently from Trilling et al. (2002), in the model of this paper the 
pile-up of planets at small periods is naturally explained.
In
our model, the distribution of internal planets is naturally connected
to the evaporation of planetesimals at very short radii.
In Trilling et al. (2002), 
the observed peak at small separations is not reproduced 
because they have not introduced any  
process to stop migration close to the central star.

The lack of a peak in Trilling's result at small separation is clearly visible in Fig 3 (light solid line), this 
last figure also shows that for $r \geq 3$ AU Trilling's simulations give a larger number of planets than the present paper.  
In other words, the model of this paper gives a better prediction, with respect to Trilling's model, of the distributon of planets observed to date both at small and large radii, while at small radi the model of this paper is probably a better description of data, at larger radii future 
surveys may help determine whether the model of this paper or Trilling's is a better description of data.
As is known, the radial velocity technique is
most sensitive to planets at small semi-major axes, because the magnitude of the stellar
wobble is greatest for close-in planets. Additionally, since the highest precision radial
velocity surveys have relatively short baselines of data, planets with longer periods are
only beginning to be identified (e.g., 55 Cnc d, with $\simeq $15 year orbit (Marcy et al. 2002)).
Thus, a large population of Jupiter-like planets could be as yet undetected.

Armitage et al. (2002) extended the work of Trilling et al.
(2002). They include a physical mechanism for disc dispersal 
into a model for the formation and migration of massive planets, 
and make a quantitative comparison with the
observed distribution of planetary orbital radii.
The results are comparable to those of Trilling et al. (2002).


Another interesting comparison is with Udry et al (2003). 
The authors showed that:\\
1) No massive planets ($M_{\rm p} \sin i \ge 2 M_{\rm J}$) are found
on short-period orbits ($P \leq 100$ days) around single stars.\\
2) The maximum mass of detected planets per period
interval increases with distance to the central star (their Fig. 6).\\
3) The migration
rate of planets decreases with increasing mass of the
planetary companion. \\
4) Up to now, no planet candidates with very low
masses ($M_{\rm p}  \sin i \leq 0.75 M_{\rm J}$) have been detected in orbits
with periods longer than $\simeq 100$ days. \\
5) There is a shortage of planets with periods in
the 10–-100 days range. 

The previous results are in agreement with the results of this paper, except, 
partially, with point 4. In this paper, low-masses planets ($M_{\rm p}  \sin i \leq 0.75 M_{\rm J}$)
tend to move towards small periods, but some can be found at periods larger than
100 days. 
This could be due to the fact that planet detectability is  poorer for increasing values of semi-major axis 
and small planet mass. If this feature is confirmed in future studies, showing a 
sharp feature in the period versus mass diagram, this could provide 
a proof of the importance of the runaway migration process recently studied by
Masset \& Papaloizou (2003). 



Our model, similarly to that of Trilling et al. (2002), 
does not deal with the distribution of orbital eccentricities of the known
radial velocity companions.
The discussion
of whether radial migration in gaseous or particulate disks produces eccentric orbits is
an ongoing one (e.g., Ward 1997a, 1997b; Bryden et al. 1999; Kley 2000; Papaloizou et
al. 2001; Murray et al. 2002). The presence of high eccentricities suggests that multiple
giant planet formation may be the norm, and that interactions among giant planets after
disk migration ends may increase the orbital eccentricities. The stochastic nature of such
interactions precludes their inclusion in an analysis such as the one we present. Further,
until the orbital inclinations can be determined, an important clue to the genesis of
the eccentricities is missing. Regardless of the mechanism(s) by which eccentricity is
created, except for extreme cases, our results regarding the semi-major axis distribution
are hardly affected \footnote{Our statistical results regarding planet formation frequency hinge on
the frequency of the close-in giant planets versus those farther out.}. Additional modest
semi-major axis evolution associated with eccentricity pumping does not alter our results
significantly. Dramatic three-planet interactions that would leave giant planets in vastly
altered orbits and greatly affect our statistics are arguably rare. In summary, while we
do not calculate eccentricities induced by gas drag in our one-dimensional model, the
results based on the statistics of semi-major axis distribution remain robust.

We recall that Woolfson (2003) showed that even in a model of planet formation that produces planets on extensive eccentric orbits, the presence of a disk-like resisting medium can explain all types of planetary orbits that have been observed, when parameters are varied in a systematic manner.

\section{Conclusions}

In this paper, we have studied the period and mass distribution of extra-solar planets, obtained 
using the migration model introduced in DP1.
The results show several features of those distributions that are in almost all cases 
in agreement with observations. 
In the case of period distribution, the paper shows:\\
1) A pile-up of planets at $a \simeq 0.05$ AU. This pile-up is clearly 
significant, and it suggests that some mechanism must be responsible for halting the migration 
of planets at an orbital period of 3 days. According to DP3 and this paper,
the distribution of internal planets is naturally connected
to the evaporation of planetesimals at very short radii.\\ 
2)  There is a  minimum near 0.3 AU, indicating a paucity of planets at that distance.\\ 
3) There is the suggestion of a rise
for semi-major axes larger than 0.3 AU, out to 3 AU. 
The increasing number of planets in larger orbits is consistent with models that invoke orbital migration and a contemporaneous clearing of the gaseous disk to explain the final positions of giant planets. 

In the case of mass distribution, we have that:\\
i) The more massive planets (typically, masses larger
than $4 M_{Jup}$) form preferentially in the outer regions 
and do not migrate much. \\
ii) Intermediate-mass objects migrate more easily
whatever the distance where they form and 
they are observed at almost all distances.\\
iii) The lighter planets (masses from sub-Saturnian to
Jovian) migrate easily. They are not confined to regions of period $< 100$ days.\\

The results obtained are in agreement with observations except item iii. In this study, we do not 
find that lighter planets are confined in regions of period $< 100$ days.
Future observations could be decisive,
showing whether mechanisms like the runaway process 
studied by Masset \& Papaloizou (2003) are needed to explain this sharp feature in the observed mass 
period diagram.

\section*{Acknowledgments}
We are grateful to P. J. Armitage for stimulating discussions during
the period in which this work was performed. A. Del Popolo would
like to thank the Bo$\breve{g}azi$\c{c}i University Research
Foundation for the financial support through the project code
01B304.

%
%

\end{document}